\documentclass{revtex4}

\usepackage{babel}
\usepackage{amsmath}
\usepackage{amssymb}
\usepackage{graphicx}
\usepackage{algorithm}
\usepackage{algorithmic}

\newcommand{\x}{\boldsymbol{x}}
\newcommand{\V}{\boldsymbol{v}}
\newcommand{\w}{\boldsymbol{w}}
\newcommand{\p}{\boldsymbol{p}}
\newcommand{\R}{\mathcal{R}}
\begin{document}

\title{Controlled Langevin Dynamics for Sampling of Feedforward Neural Networks Trained with Minibatches}

\author{Alessandro Zambon$^1$, Francesca Caruso$^2$, Riccardo Zecchina$^{2*}$ and Guido Tiana$^{1*}$ }

\affiliation {$^1$Department of Physics, Università degli Studi di Milano and INFN, via Celoria 16, 20133 Milano, Italy}
\affiliation{$^2$Department of Computing Sciences and Bocconi Institute for Data Science and Analytics (BIDSA), Bocconi University, 20136 Milano, Italy}

\affiliation{$^*$Authors to whom any correspondence should be addressed.}

\email{riccardo.zecchina@unibocconi.it, guido.tiana@unimi.it}


\begin{abstract}
Sampling the parameter space of artificial neural networks according to a Boltzmann distribution provides insight into the geometry of low–loss solutions and offers an alternative to conventional loss minimization for training. However, exact sampling methods such as hybrid Monte Carlo (hMC), while formally correct, become computationally prohibitive for realistic datasets because they require repeated evaluation of full-batch gradients. We introduce a pseudo-Langevin (pL) dynamics that enables efficient Boltzmann sampling of feed–forward neural networks trained with large datasets by using mini--batches in a controlled manner. The method exploits the statistical properties of mini--batch gradient noise and adjusts fictitious masses and friction coefficients to ensure that the induced stochastic process samples efficiently the desired equilibrium distribution. We validate numerically the approach by comparing its equilibrium statistics with those obtained from exact hMC sampling. Performance benchmarks demonstrate that, while hMC rapidly becomes inefficient as network size increases, the pL scheme maintains high computational diffusion and scales favorably to networks with over one million parameters. Finally, we show that sampling at intermediate temperatures yields optimal generalization performance, comparable to SGD, without requiring a validation set or early stopping procedure. These results establish controlled mini--batch Langevin dynamics as a practical and scalable tool for exploring and exploiting the solution space of large neural networks.
\end{abstract}

\maketitle

\section{Introduction}

The parameters that enable an artificial neural network (ANN) to make predictions are usually obtained by minimizing a loss function with a stochastic gradient descent (SGD) algorithm. However, exploring the low-loss manifold at large has proven useful for finding parameters with better generalization properties and for understanding the fundamental principles underlying the operation of the network. 
For example, sampling the space of parameters within a Monte Carlo algorithm driven by molecular dynamics moves within the framework of the canonical ensemble, elucidated that low--loss parameters are arranged according to a spiky topology  close to the interpolation threshold in a small tree committee machine, while it becomes flat in the over--parametrized regime \cite{Zambon2025SamplingNetwork}. 
A replica algorithm allowed the identification of wide minima of the loss function, which were shown to generalize more effectively than those identified using SGD \cite{Baldassi2015,Baldassi2016UnreasonableSchemes,Pittorino2020}. 
Analogously, Bayesian learning, which requires sampling of the parameter space of the ANN, shows several advantages with respect to standard training methods \cite{MacKay1992ANetworks}, such as improved generalization capacity \cite{Welling2011BayesianDynamics}.
All these results suggest that sampling the solution space could be an alternative to straightforward loss minimization through SGD, even for the routine training of an ANN.

The main problem when sampling the parameter space of neural networks is the efficiency of the sampling algorithm. The hybrid Monte Carlo method employed in \cite{Zambon2025SamplingNetwork} has the virtue of sampling according to the desired Boltzmann distribution, while the elementary steps are guided by gradient-based molecular dynamics, which are critical to navigate such a high-dimensional space. However, this algorithm is difficult to use with training sets of a realistic size because calculating the gradient becomes lengthy. When the task is just loss minimization, the use of mini--batches tries to solve this problem; however, mini--batches cannot be applied in a straightforward manner to the Monte Carlo scheme as they would violate the detailed balance requirement.

In this work, we introduce an efficient sampling scheme to sample the space of solutions of a feed--forward ANN of realistic size, approximating the Boltzmann distribution. This method is based on a Langevin--like equation and uses mini--batches to make the exploration efficient even in presence of large training sets. The Langevin equation defines a stochastic process in the space of parameters that converges to the Boltzmann distribution \cite{VanKampen2003}. This kind of approach has been used to sample the posterior probability of the parameters in Bayesian training of ANN, using mini--batches in a naive way, which essentially assumes that the error introduced by its use is negligible  \cite{Li2015}. In another study using Langevin dynamics, the effect of the noise introduced by mini--batches was explicitly calculated and subtracted from the dynamics at each step \cite{Chen2014}. However, since this requires the explicit calculation of the covariance matrix between the components of the gradient averaged over different mini--batches, it becomes unfeasible for large datasets.

The algorithm we developed samples efficiently the space of parameters of an ANN according to Boltzmann's weight through a control of the noise generated by the estimation of the loss function with mini--batches. We compared the results obtained with this algorithm with those coming from a hybrid Monte Carlo sampling, which is asymptotically correct but inefficient, basically useless for large networks trained with datasets of realistic size. Moreover, we suggest that the sampling could be used as an efficient alternative to SGD for training large networks, a procedure that offers some advantages for applications.

\section{Methods}

The following sections provide descriptions of the models used to validate the sampling method and the hybrid Monte Carlo algorithm. The latter provides asymptotically correct sampling, but is inefficient for large datasets. Lastly, we describe our Langevin strategy based on mini-batches.

    \subsection{The model and task}
    \label{sec:model_and_task}

        \begin{figure}
    		\centerline{\includegraphics[width=0.3\linewidth]{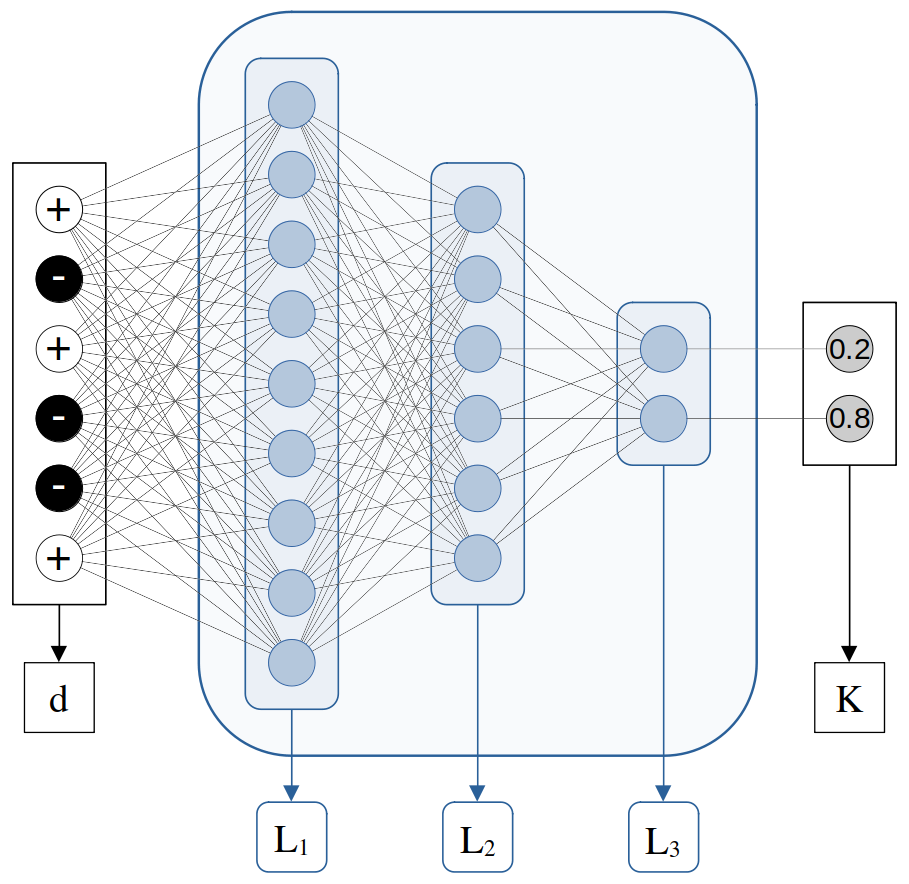}}
    		\caption{Sketch of the feed--forward ANN used in this work, with (from left to right) $L_1$, $L_2$ and $L_3$ neurons for the first, second and third layer respectively. On the left and on the right of the NN, a possible couple of input and output vectors respectively, that is a $d$-dimensional spin vector and a normalized probability distribution on the $K=L_3$ classes.}
    		\label{fig:FFN}
    	\end{figure}

        We studied a standard feed--forward ANN, made of three fully-connected layers of size $L_1$, $L_2$ and $L_3$, respectively, with ReLU activation functions for all neurons (Fig.~\ref{fig:FFN}). This can be described as
        \begin{equation}
    		\begin{split}
                \boldsymbol{a}^{(l)} &= W^{(l , \, l-1)} \cdot \boldsymbol{z}^{(l-1)} + \boldsymbol{b}^{(l)} \\
                \boldsymbol{z}^{(l)} &= \text{max}[0, \boldsymbol{a}^{(l)}]
    		\end{split}
    		\label{eq:FFN}
    	\end{equation}
        with $W^{(l , \, l-1)} \in \mathbb{R}^{L_l \times L_{l-1} }$ and $\boldsymbol{b}^{(l)} \in \mathbb{R}^{L_l}$ (for $l=1,2,3$) are the learnable weights of the network. In addition, $\boldsymbol{z}^{(0)} \equiv \x$ is the input vector, while $\hat{\boldsymbol{y}} = \text{softmax}(\boldsymbol{z}^{(L_3)})$ is the output probability distribution. We will refer to the entire set of parameters of the network as the weight vector $\w$.

        The task we want to address is a classification of groups of random vectors, which are similar to each other in each class. We consider $K$ classes, each associated with a reference vector $\V^{(k)} \in \{ \pm 1 \}^{d}$ of dimension $d$, extracted from a flat distribution, that is $v^{(k)}_i \sim \frac{1}{2} \left( \delta_{v^{(k)}_i, +1} + \delta_{v^{(k)}_i, -1} \right)$ for $k=0,...,K-1$ and $i=0,...,d-1$. We then proceed to the generation of the dataset $\mathcal{D} = \{(\x^\mu, y^\mu)\}_{\mu=1}^{P}$, made of random vectors that are similar to each of the representatives of the classes. For each example $\mu=0,...,P-1$,
        \begin{equation}
            \begin{split}
                y^\mu &\sim K^{-1} \sum_{k=1}^{K} \delta_{y^\mu, k} \\
                x^\mu_i &\sim (1-p_{f}) \, \delta_{x^\mu_i, v^{(y^\mu)}_i} + p_{f} \, \delta_{x^\mu_i, -v^{(y^\mu)}_i} \qquad \forall i=0,...,d-1,
            \end{split}
            \label{eq:dataset}
        \end{equation}
        where $p_f$ is the probability of flipping (that is, changing the sign of) each component of the $\mu$-th example reference vector $\V^{(y^\mu)}$. In order to adapt the ANN presented in Eq~\eqref{eq:FFN} to the current task, we will keep $L_2=d$ and $L_3=K$ fixed, while the dimension of the network $N$ (i.e. the size of the weight vector $\w$) is  determined by the number of neurons in the first hidden layer $L_1$.

        Given the task at hand, we characterize each weight vector with a potential function composed of two terms, namely the mean cross-entropy loss function and a $L^2$ regularization term,
        \begin{equation}
            U\left(\w | \mathcal{D}\right) = -\frac{1}{P} \sum_{\mu=1}^{P} \sum_{k=1}^{K}\delta_{y^\mu, k} \, \text{ln}\;\hat{y}^\mu_k + \frac{\lambda}{2N} \sum_{i=1}^{N} w_i^2.
            \label{eq:potential}
        \end{equation}
        In this study, we  keep the ratio $\alpha=P/N$, called 'constrained density', fixed while increasing the size of the network $N$. For this reason, we considered an effective Lagrange multiplier $\lambda/N$, to make both the loss function and the regularization term independent of the scale of the system. 
        Furthermore, another observable we can use to characterize each weight vector $\w$ is the fraction of misclassified input-output couples, that is,
        \begin{equation}
            \epsilon\left(\w | \mathcal{D}\right) = \frac{1}{P} \sum_{\mu=1}^{P} 1-\delta_{y^\mu, \hat{k}^{\mu}},
            \label{eq:errorfunction}
        \end{equation}
        where $\hat{k}^{\mu} = \mathrm{argmax}_{k} \, \hat{y}^{\mu}_{k}$ is the most-probable predicted class.
        

    \subsection{The hybrid Monte Carlo}
        \label{sec:hMC}
        A way to sample the Boltzmann distribution is a Monte Carlo algorithm, as the hybrid (hMC) Monte Carlo used in ref. \cite{Zambon2025SamplingNetwork}, which is exact in the limit of a large number of moves.

        Operatively, at each iteration $j$, the momenta $\p_j$ associated with the current weight vector $\w_j$ are extracted from a Maxwell--Boltzmann distribution $\mathcal{N}(0,\mathcal{M}T)$, where $\mathcal{M}$ is a diagonal matrix of (fictitious)  masses and $T$ is the temperature in the weight space (we also define $\beta \equiv T^{-1}$). Next, we employ a velocity Verlet algorithm to calculate an approximate solution of the equations of motion associated with the Hamiltonian
        \begin{equation}
            H(\w, \p) = \frac{1}{2} \p^{T} \mathcal{M}^{-1} \p + U(\w), 
            \label{eq:Hamiltonian}
        \end{equation}
        for a total time $\Delta t$, using a finite integration time step $\delta t$, where the initial conditions are $\left( \w(0), \p(0) \right) \equiv \left( \w_j, \p_j \right)$.
        Since the total energy is generally not conserved exactly when using a finite integration step, at the end of the trajectory the last weight vector $\w(\Delta t)$ is accepted as the new starting configuration with probability $\min(1,\exp[-\Delta H/T])$, where $\Delta H = H[\w(\Delta t), \p(\Delta t)] - H[\w(0), \p(0)]$ is the total energy variation. If accepted, then the process is repeated from $\w_{j+1} = \w(\Delta t)$, otherwise we set $\w_{j+1} = \w(0) = \w_j$. Due to the invariance for time inversion of the integration algorithm ($\delta t\to -\delta t$), the principle of detailed balance is satisfied and the algorithm samples exactly Boltzmann's distribution.

        The strength of this algorithm is that the energy gradient chooses the optimal direction for the proposed move, which is critical in such a high--dimensional space. However, this is also its main drawback since the calculation of the gradient has a large computational cost. 
        
        One way to reduce this cost would be to evaluate the gradient on mini--batches, that is subsets of $\mathcal{D}$ selected at random at the beginning of each hMC step. The acceptance procedure then would be carried out with the exact energy calculated on the whole $\mathcal{D}$ to guarantee the sampling of the correct distribution. Unfortunately this algorithm, which we call 'mb-hMC', is still inefficient because it requires a very small $\Delta t$ to prevent the Metropolis acceptance probability from becoming negligible (cf. Fig. \ref{app_fig:othermethods_performance}).

    \subsection{A pseudo-Langevin (pL) algorithm}
        \label{sec:PL}
        Another way of sampling the canonical ensemble is through Langevin dynamics, which is described by the stochastic differential equations
        \begin{equation}
    		\begin{split}
                d\w(t) &= \mathcal{M}^{-1}\p(t)dt \\
                d\p(t) &= - \gamma\p(t)dt -\nabla U \left(\w(t) \right) dt + \sqrt{2\mathcal{M}\gamma T} d\boldsymbol{W}(t),
    		\end{split}
    		\label{eq:SDELangevin}
    	\end{equation}
        where $\mathcal{M}$ is once again a diagonal mass matrix, $\gamma > 0$ is the friction coefficient and $dW_i(t)$ is a Wiener process with $\langle dW_i(t) dW_j(t')\rangle = \delta_{i,j}\delta(t-t')$ for $i,j=0,...,N-1$. An efficient way to solve Eqs. (\ref{eq:SDELangevin}) is to use the Bussi--Parrinello integrator \cite{Bussi2007a}
        \begin{equation}
    		\begin{split}
                \w(t + \delta t) &= \w(t) + \mathcal{M}^{-1}\boldsymbol{\Pi}(t)\delta t\\
                \boldsymbol{\Pi}(t + \delta t) &= c_{1}^{2} \boldsymbol{\Pi}(t) - \frac{(1+c_1^2)\delta t}{2}\nabla U \left( \w(t + \delta t) \right) + \sqrt{1+c_1^2}C_2\boldsymbol{\R}(t+\delta t),
    		\end{split}
    		\label{eq:BussiParrinello}
    	\end{equation}
        where $\R_{i}(t) \sim \mathcal{N}(0, 1)$ are extracted from independent Gaussian distributions, $\delta t$ is the integration time step, $c_1 = e^{-\gamma \delta t / 2}$ accounts for the dissipation caused by the viscosity of the medium, $C_2^2 = (1-c_{1}^{2})\mathcal{M}T$ is the variance matrix of the applied white noise at temperature $T$ and  $\boldsymbol{\Pi}(t)$ is a short-hand for 
        \begin{equation}
    		\boldsymbol{\Pi}(t) = c_1\p(t) - \frac{\delta t}{2}\nabla U \left( \w(t) \right) + C_2 \boldsymbol{\R}(t).
    		\label{eq:ExtendedMomenta_BP}
    	\end{equation}

        
        However, this method still requires the computation of the energy gradient, so it suffers from the same computational inefficiency as the hMC. To address this issue, we evaluated the gradient on mini--batches in a controlled manner. 
        We define the set of example indexes of the dataset $\mathcal{D}$ as $\mathcal{I}=\{1, 2, ...,P\}$. For each given weight vector $\w$, we can associate with $\mathcal{I}$ a set $\mathcal{G}_i(\w)$, containing the values assumed by the $i$-th component of the energy gradient computed over all the examples in the dataset, that is
        \begin{equation}
            \mathcal{G}_i(\w) = \{ \nabla_i U^{(\mu)}(\w)\}_{\mu \in \mathcal{I}},
            \label{eq:SetOfGradients}
        \end{equation}
        where $U^{(\mu)} \left(\w\right)$ is the potential function for a weight vector $\w$ evaluated on the $\mu$-th example.

        We now promote the example index $\mu$ to a stochastic variable extracted from a flat distribution over the set $\mathcal{I}$,
        \begin{equation}
            \mu \sim \textnormal{Unifom}\left( \mathcal{I} \right), 
        \end{equation}
        thus obtaining an estimate for the elements of the gradient on a single example
        \begin{equation}
            \nabla_i U^{(\mu)}(\w) \sim \textnormal{Unifom}\left( \mathcal{G}_i (\w) \right).
            \label{eq:ExampleAndGradientExtraction}
        \end{equation}
        Such an estimate satisfies 
        \begin{equation}
                \langle \nabla_i U^{(\mu)} \left( \w \right) \rangle_{\mu} = \nabla_i U \left( \w \right), 
                \label{eq:GradientProperties1}
        \end{equation}
        because the function in Eq~\eqref{eq:potential} is intensive with respect to the number of examples $P$ and also
        \begin{equation}
                \textnormal{Var} \left[ \nabla_i U^{(\mu)} \left( \w \right) \right] < \infty,
            \label{eq:GradientProperties2}
        \end{equation}
        since the set $\mathcal{G}_i (\w)$ is finite and limited under the realistic assumption that $U$ is differentiable everywhere.

        A mini-batch $b$ of size $S$ can be regarded as a stochastic process $b = \{ \mu_n | \mu_n \in \mathcal{I} \}_{n=1}^{S}$ and, analogously, the $i$-th component of the energy gradient evaluated on such mini-batch is the mean of the other associated stochastic process $\{ \nabla_i U^{(\mu_n)} \left( \w \right) | \mu_n \in \mathcal{I} \}_{n=1}^{S}$, that is,
        \begin{equation}
            \nabla_i U^{(b)} \left( \w \right) = S^{-1} \sum_{n = 1}^{S} \nabla_i U^{(\mu_n)} \left( \w \right).
            \label{eq:MiniBatchGradientDefinition}
        \end{equation}
         
        Although the set $\mathcal{G}_i(\w)$ is generally composed of correlated values, the stochastic variables $\nabla_i U^{(\mu_n)} \left( \w \right)$ are extracted independently from the same distribution $\textnormal{Unifom}\left( \mathcal{G}_i (\w) \right)$. Together with the properties reported in Eqs. \eqref{eq:GradientProperties1} and \eqref{eq:GradientProperties2}, this justifies the application of the central limit theorem for $S \gg 1$; the $i$-th component of the mini--batch gradient thus satisfies
        \begin{equation}
            \nabla_i U^{(b)} \left( \w \right) \sim \mathcal{N} \left( \nabla_i U \left( \w \right),  \frac{\textnormal{Var} \left[ \nabla_i U^{(\mu)} \left( \w \right) \right]}{S}\right).
            \label{eq:MiniBatchGradientDistribution}
        \end{equation}
        The result from Eq~\eqref{eq:MiniBatchGradientDistribution} is valid for each component $i$ and each weight vector $\w$, thus we can easily extend it to the vector $\nabla U^{(b)}$ by defining $\mathcal{V}(\w|S)$ as a diagonal matrix of the variances associated with each component of the mini-batch gradient. However, since the latter are not independent in principle, we also introduce the correlation matrix between the gradient components $\mathcal{C}(\w|S)$, such that $\mathcal{C}_{ii}(\w|S) = 0 \,\, \forall i$. Therefore, the gradient of the potential energy calculated on a stochastic mini-batch $b$ is distributed as follows,
        \begin{equation}
            \nabla U^{(b)} \left( \w \right) \sim \mathcal{N} \left( \nabla U(\w), \mathcal{V}(\w|S)+\mathcal{C}(\w|S) \right).
            \label{eq:MiniBatchGradientGaussianity}
        \end{equation}
        We can now approximate $\nabla U \left( \w(t) \right)$ with $\nabla U^{(b)} \left( \w(t) \right)$ to guide the sampling, by substituting
        \begin{equation}
            \frac{\delta t}{2} \nabla U \left(\w\right)  + \mathcal{N}(0, C_2^2) = 
            \frac{\delta t}{2} \nabla U^{(b)} \left(\w\right) + \mathcal{N} \left[ 0, C_2^2 - \left( \frac{\delta t}{2} \right)^2 \left[ \mathcal{V}\left(\w\right) + \mathcal{C}\left(\w\right) \right] \right].
    		\label{eq:substitution}
    	\end{equation}
        
        The calculation can be made even more efficient. First of all, the estimation of the non-diagonal elements of mini-batch noise covariance matrix (i.e. $\mathcal{C}(\w)$) can be  computationally expensive for increasing $N$. Therefore, we can adjust the masses $\mathcal{M}$ (which are fictitious, so we are free to choose them) in such a way that
        \begin{equation}
            \mathcal{T}_i \equiv \left( \frac{\delta t}{2} \right)^2 \left[ \left( C_2^2 \right)^{-1} \mathcal{V}\left(\w\right) \right]_{ii} = \frac{\delta t^2}{4(1-c_1^2)T} \left[ \mathcal{M}^{-1} \mathcal{V}_{\tau} \right]_{ii}
            \ll 1 \quad \forall i
            \label{eq:LTRLimit}
        \end{equation}
        i.e. the error introduced by computing $\nabla U^{(b)} \left(\w\right)$ instead of $\nabla U \left(\w\right)$ is small compared to the order of magnitude of the total expected noise. We shall refer to this quantity $\mathcal{T}_{i}$ as 'temperatures ratio', and the condition expressed in Eq.\eqref{eq:LTRLimit} as the 'low temperatures ratio' (LTR) limit.
        This condition, along with the Cauchy--Schwarz inequality
        \begin{equation}
            |\mathcal{C}_{ij} \left(\w\right)| \leq \sqrt{ \mathcal{V}_{ii} \left(\w\right) \mathcal{V}_{jj} \left(\w\right) } \ll \sqrt{ \left[ \left( \frac{\delta t}{2} \right)^{-2}C_2^2 - \mathcal{V}\left(\w\right) \right]_{ii} \left[ \left( \frac{\delta t}{2} \right)^{-2}C_2^2 - \mathcal{V}\left(\w\right) \right]_{jj}},
    		\label{eq:remove_correlation}
    	\end{equation}
        allows us to approximate the expression in Eq~\eqref{eq:substitution} to
        \begin{equation}
            \frac{\delta t}{2} \nabla U \left(\w\right)  + \mathcal{N}(0, C_2^2) \approx 
            \frac{\delta t}{2} \nabla U^{(b)} \left(\w\right) + \mathcal{N} \left[ 0, C_2^2 - \left( \frac{\delta t}{2} \right)^2 \mathcal{V}\left(\w\right) \right],
    		\label{eq:ApproximationLTR}
    	\end{equation}
        which does not require the calculation of either $\nabla U(\w)$ or $\mathcal{C}(\w)$. Of course, this simplification comes at the cost of checking periodically that the condition of Eq~\eqref{eq:LTRLimit} is verified along the trajectory.

        Moreover, we can remove the slowness associated with the fact that in principle the matrix $\mathcal{V}(\w)$ changes along with the current weight vector and thus should be continuously updated. To do so, we regard the noise variances of mini--batches as constants for short periods during the trajectory,
        \begin{equation}
            \mathcal{V}(\w(t)) \approx \mathcal{V}(\w(\tau)) \equiv \mathcal{V}_{\tau} \quad \forall t \in \left[ \tau, \tau+\Delta \right),
            \label{eq:ApproximationConstance}
        \end{equation}
        where $\Delta$ is the period after which the expected variances of the noise $\mathcal{V}_{\tau}$ are updated. The reason for this approximation is that the mini-batch noise is a small fraction of the total noise injected into the system (see Eq~\eqref{eq:LTRLimit}). Therefore, temporarily neglecting fluctuations of the order $\mathcal{O}\left( \mathcal{V}_\tau \right)$ should not have a significant impact on the sampling.
        
        In Sect. \ref{sec:validation}, we shall analyze the properties of the 'mini--batch noise', defined as 
        \begin{equation}
            \boldsymbol{\R}^{(b)}_{\tau}\left(\w(t)\right) \equiv \mathcal{V}_{\tau}^{-\frac{1}{2}} \left[ \nabla U^{(b)}\left(\w(t)\right)-\nabla U\left(\w(t)\right) \right],
            \label{eq:MiniBatchNoise}
        \end{equation}
        that is the difference between the gradient computed on a stochastic mini--batch and on the whole dataset, scaled by the last updated standard deviation matrix, and we will verify that the approximations made in Eq.~\eqref{eq:ApproximationLTR} and Eq.~\eqref{eq:ApproximationConstance} are verified along the dynamics.

        Taking advantage of the Gaussian character of the noise generated by mini-batches under the  simplifications of Eq.~\eqref{eq:ApproximationLTR} and Eq.~\eqref{eq:ApproximationConstance}, we obtain the pseudo-Langevin (pL) integrator,
        \begin{equation}
    		\begin{split}
                \w(t + \delta t) &= \w(t) + \mathcal{M}^{-1}\boldsymbol{\Pi}(t)\delta t\\                
                \boldsymbol{\Pi}(t + \delta t) &= c_{1}^{2} \boldsymbol{\Pi}(t) - \frac{(1+c_1^2)\delta t}{2} \nabla U^{(b_{t+\delta t})} \left( \w(t + \delta t) \right) + \\
                &+ \sqrt{(1+c_1^2) \left[ K_{\tau}^{2} - \left( \frac{c_1\delta t}{2} \right)^2 \mathcal{V}_{\tau} \right]} \boldsymbol{\R}(t+\delta t),
            \end{split}
            \label{eq:PseudoLangevinIntegrator}
        \end{equation}
        where $b_t$ is the mini-batch selected at time $t$, $K_{\tau}^2 \equiv C_2^2 - \left( \frac{\delta t}{2} \right)^2 \mathcal{V}_{\tau}$ is the variance matrix of the white noise added on top of the correlated one associated with the mini-batch $b_t$ and, similarly to Eq.~\eqref{eq:ExtendedMomenta_BP},
        \begin{equation}
            \boldsymbol{\Pi}(t) = c_1 \p(t) - \frac{\delta t}{2}\nabla U^{(b_t)} \left( \w(t) \right) + K_{\tau} \boldsymbol{\R}(t).
            \label{eq:ExtendedMomenta_pL}
        \end{equation}
        One should note that Eqs.~\eqref{eq:PseudoLangevinIntegrator} is valid only if
        \begin{equation}
    		K_{\tau}^{2} - \left( \frac{c_1\delta t}{2} \right)^2 \mathcal{V}_{\tau} \geq 0 \Rightarrow \mathcal{T}_i  \leq \frac{1}{1+c_1^2} \quad \forall i=0,...,N-1,
            \label{eq:ConditionForIntegrability}
        \end{equation}
        meaning that, in general, $\mathcal{T}_i \leq 0.5$ (since $c_1 \in (0, 1)$). Thus, this condition is actually already included in the LTR condition of Eq. (\ref{eq:LTRLimit}). 

\begin{figure}
    		\centerline{\includegraphics[width=1.0\linewidth]{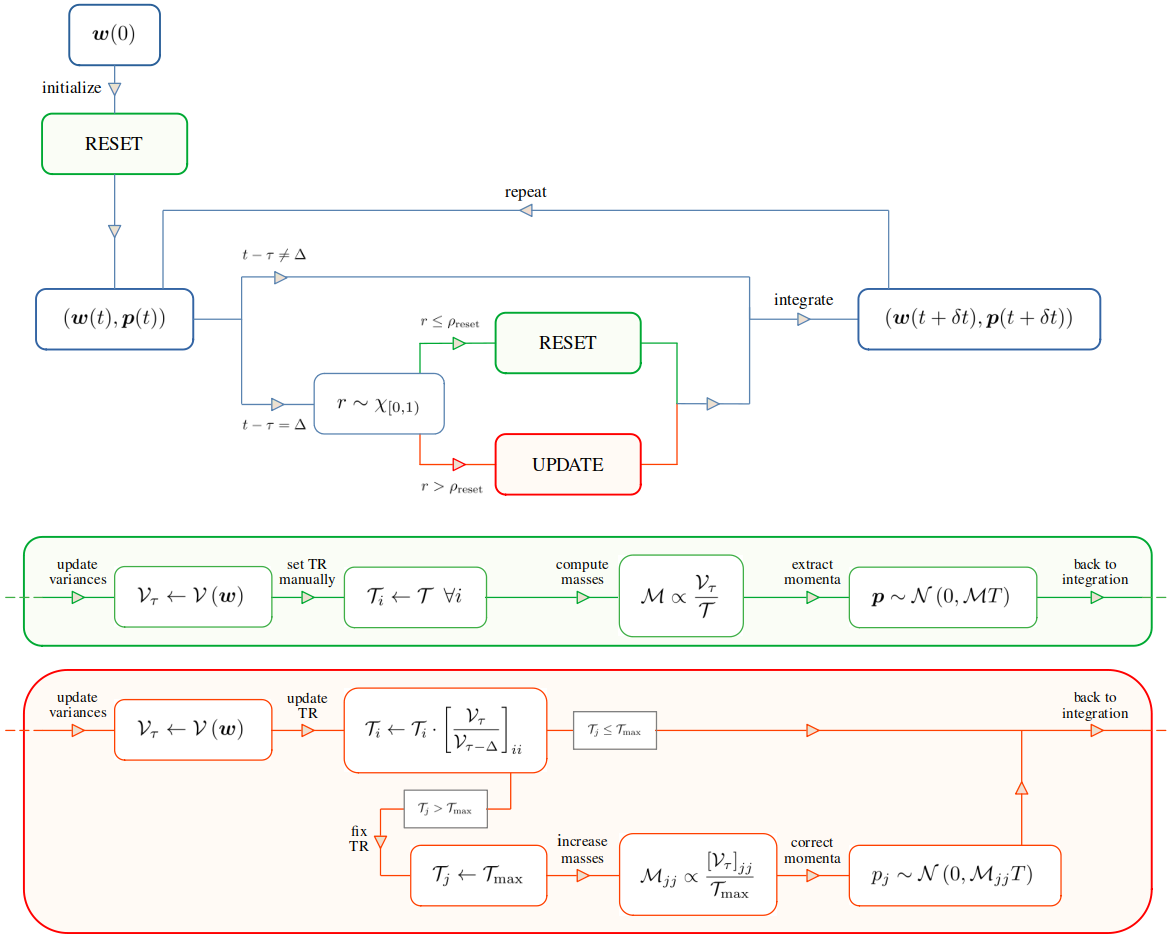}}
    		\caption{Scheme of the pseudo--Langevin (pL) algorithm.}
    		\label{fig:pL_scheme}
        \end{figure}

        In practice, at time $t_0$, the matrix $\mathcal{V}_{\tau}$ is initialized (with $\tau=t_0$) by extracting mini-batches of fixed size $S$ until the estimate of the variance converges for each gradient component. Next, the initial values of the temperature ratios, $\mathcal{T}=\mathcal{T}_i$ (i.e. the same for each weight component), and the dissipative term, $c_1$, are set manually. The mass matrix is then calculated according to Eq.~\eqref{eq:LTRLimit} using $\delta t=1$. The momenta are then extracted from the Maxwell–Boltzmann distribution $\mathcal{N}(0,\mathcal{M}T)$ and $K_{\tau}$ is computed. Once the initialization procedure has ended, a trajectory in phase space is generated using Eq.~\eqref{eq:PseudoLangevinIntegrator} and the mini-batch noise variance matrix, $\mathcal{V}_{\tau}$, is periodically updated using the same procedure as at time $t_0$. During each update, the current temperature ratios are measured and the LTR condition is checked using a threshold value, i.e. a maximum temperature ratio, $\mathcal{T}_{\mathrm{max}}$, such that $\mathcal{T}_i \leq \mathcal{T}_{\mathrm{max}} \,\, \forall i$. If the previous inequality is not satisfied for certain components, their masses are increased until their associated temperatures ratios are equal to the threshold value. The corresponding momenta are then extracted from the correct Maxwell–Boltzmann distribution once again. Lastly, the initialization procedure can be carried out with probability $\rho_{\mathrm{reset}}$ during each update, resetting the value of the masses and extracting the momenta from the true equilibrium distribution. The algorithm pipeline is presented schematically in Fig.~\ref{fig:pL_scheme}.

        A key point is that, in principle, the Bussi--Parrinello integrator [Eq. ~\eqref{eq:BussiParrinello}] samples asymptotically the correct equilibrium distribution only in the limit $\delta t \rightarrow 0$. Thus, since the effective mini--batch variance scales quadratically with $\delta t$ (see Eq~\eqref{eq:substitution}), one could manually set the values of the mass matrix $\mathcal{M}$ to some arbitrary value and employ a very small integration time-step \cite{Li2015,Chen2014}. However, this choice does not take into account that the different components of the weight vector contribute differently to the total expected noise at a fixed temperature. A wiser choice is based on the observation that Eq. (\ref{eq:BussiParrinello}) is a good approximation of the stochastic differential equation (\ref{eq:SDELangevin}) also if, given a finite step--size $\delta t$, the fictitious masses and the fictitious friction coefficient satisfy
        \begin{equation}
            \begin{split}
    		      \frac{\delta t}{\sqrt{\mathcal{M}_{ii}}} &\rightarrow 0 \,\,\, \forall i,\\
                \gamma \delta t &\rightarrow 0.
            \end{split}
            \label{eq:CorrectSamplingLimit}
        \end{equation}
        In terms of of Eq.~\eqref{eq:PseudoLangevinIntegrator}, these conditions can be expressed as 
        \begin{equation}
            \begin{split}
    		      \mathcal{T}_i &\rightarrow 0 \,\,\, \forall i, \\
                c_1 &\rightarrow 1,
            \end{split}
            \label{eq:CorrectSamplingLimit}
        \end{equation}
        meaning that the limit from Eq.~\eqref{eq:LTRLimit} not only guarantees the removal of the correlation between the components of the total noise, but also makes the dynamics produced by Langevin equations to sample the correct equilibrium distribution (for a high--enough viscosity coefficient $c_1$). In Fig. \ref{app_fig:othermethods_performance}, we show that a pL sampling method which does not use the LTR limit and that instead implements a uniformly vanishing step--size $\delta t$ (which we will call 'naive-pL') is sub-optimal with respect to pL in terms of sampling efficiency.

\section{Results}
    \label{sec:results}
    We studied three feed--forward ANN models (Fig.~\ref{fig:FFN}) of size $N=11160$, $N=101610$ and $N=1006110$, respectively. For each size, we generated a dataset of random spin vectors of dimension $d=100$, divided in $K=10$ classes (cf. Sect. \ref{sec:model_and_task}). To make the classification problem non--trivial, we chose a spin-flip probability of $p_f=0.355$, so that the fraction of spin vectors closer to the reference vector of another class than to their own is approximately $0.1$. We studied each model near the interpolation threshold, with a training dataset composed of $P=5N$ examples. Lastly, the norm of the weights is controlled by the Lagrange multiplier $\lambda\approx10$ [see Eq.~\eqref{eq:potential}], to avoid that it diffuses in an uncontrolled way.

    \subsection{Numerical validation of the pL method}
        \label{sec:validation}
        \begin{figure}
    		\centerline{\includegraphics[width=0.7\linewidth]{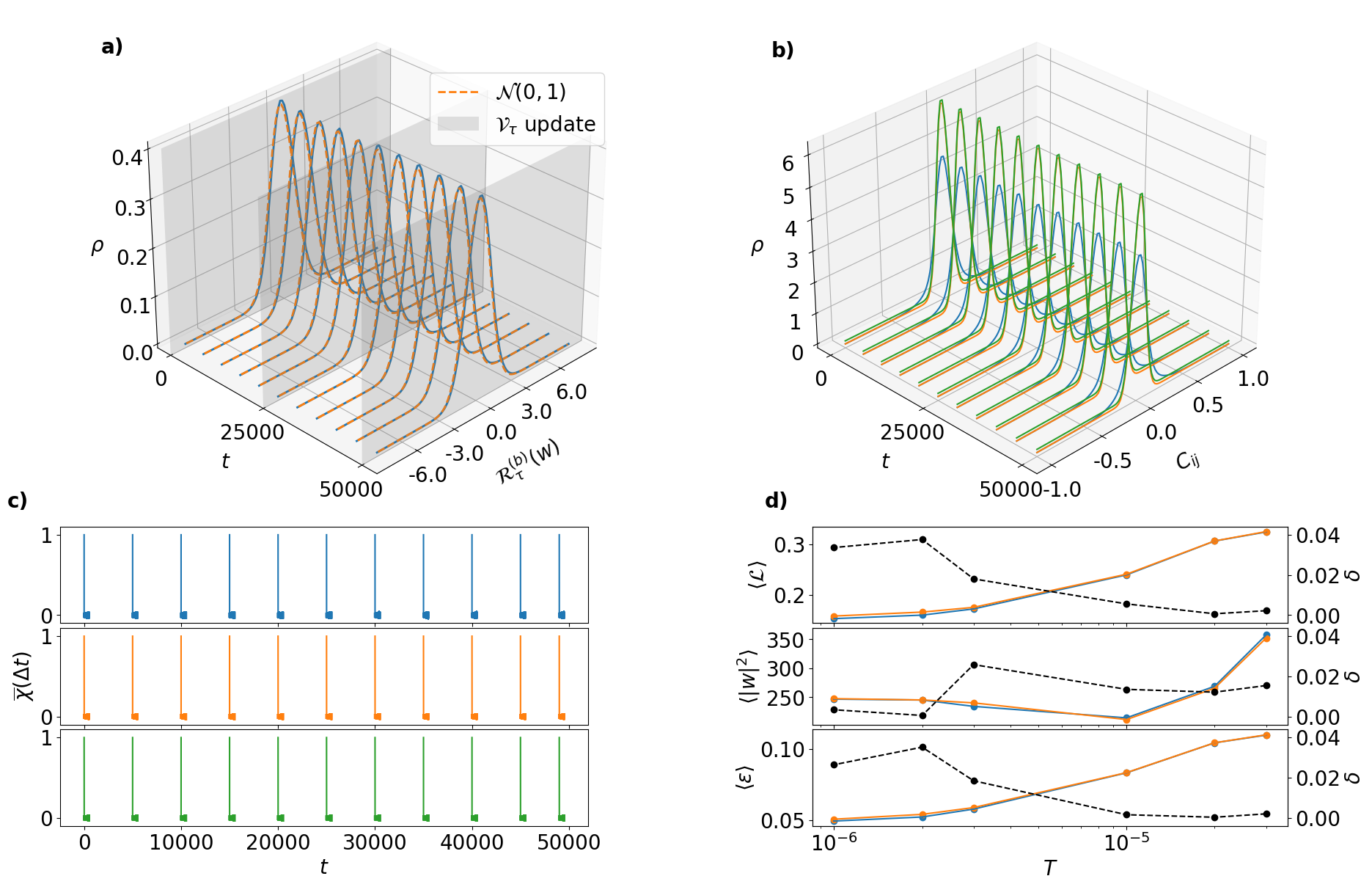}}
    		\caption{\textbf{a)} The distribution $\rho$ of the components of the mini-batch noise $\boldsymbol{\R}^{(b)}_{\tau} \left(\w(t)\right)$ during a pL simulation as a function of time $t$ (blue curves), compared to expected distribution $\mathcal{N}(0,1)$  (orange curves). The gray planes indicate the times at which the values of the mini-batch noise variance matrix $\mathcal{V}_{\tau}$ are updated.
            \textbf{b)} The distribution $\rho$ of the elements of the correlation matrix $C_{ij}$ during a pL simulation as a function of time $t$, computed among the components of the mini-batch noise $\boldsymbol{\R}^{(b)}_{\tau} \left(\w(t)\right)$ (blue curves), of the white noise $\boldsymbol{\R}(t)$ (orange curves) and of the weighted sum of the two (green curves). The curves are slightly shifted along the z-axis to make them distinguishable.
            \textbf{c)} The average of the component--wise noise autocorrelation $\overline{\chi}(\Delta t)$ for the mini-batch noise (upper panel, blue curves), for the white noise (middle panel, orange curves) and for their weighted sum (lower panel, green curves), computed for small time intervals during a pL simulation parametrized by the time $t$.
            \textbf{d)} The average of the mean cross-entropy $\mathcal{L}$ (upper plot), the squared norm $|\w|^2$ (center plot) and the training error $\epsilon$ (lower plot) as a function of the temperature $T$ obtained from hMC (blue curves) and pL (orange curves) for the smallest model $N=11160$. The relative error between the estimates is also reported (gray dotted line).}
    		\label{fig:validation}
        \end{figure}

        We first used the the smallest architecture ($N=11160$) to validate the approximations described in Sect. \ref{sec:PL}. For this purpose, we performed three independent short pL simulations  at temperature $T=10^{-6}$, mini-batch size $S = 10^{-2}P$ and temperatures ratio $\mathcal{T}=3\cdot10^{-2}$. During each simulation, we periodically stopped the integration process to calculate the full-batch gradient and extract $250$ mini-batch gradients $\{ \boldsymbol{\R}^{(b)}_{\tau}\left(\w(t)\right) \}$. Each simulation starts from a configuration obtained minimizing the loss using Adam as optimizer for $5000$ epochs with mini-batch size $s=128$ and initial learning rate $\eta=10^{-4}$ (scaled by a factor of $\gamma=0.8$ every $1000$ epochs). 

        We observe that indeed the distribution of the noise generated by the use of mini--batches on the gradient follows a Gaussian distribution and that the true variance matrix $\mathcal{V}\left(\w(t)\right)$ along the trajectory can be approximated, for sufficiently small time periods, with the last updated $\mathcal{V}_{\tau}$ (Fig.~\ref{fig:validation}a). At each time $t$ and for each gradient component $i$, we performed a Kolmogorov--Smirnov (KS) test on the distribution of extracted mini--batch noises $\{ \boldsymbol{\R}^{(b)}_{\tau} \left( \w(t) \right) \}$ (see Eq.~\eqref{eq:MiniBatchNoise}) to the standard normal distribution $\mathcal{N}(0,1)$. A fraction of $0.946$ of the tests (averaged over $t$ and $i$) displays a p-value $>0.05$, which is a result comparable to what one would get if the KS tests were performed on truly Gaussian-distributed random variables.

        Secondly, the LTR condition stated in Eq. (\ref{eq:LTRLimit}) is sufficient to eliminate the correlation between the components of the total noise produced during integration. We  calculated periodically the correlation matrix $C_{ij}$ between the components of the  mini--batch noise $\{ \boldsymbol{\R}^{(b)}_{\tau} \left(\w(t)\right) \}$ and the weighted sum of the latter with white noises $\{ \boldsymbol{\R}(t) \}$, that is $\{ \frac{\delta t}{2} \mathcal{V}^{\frac{1}{2}}_{\tau}\boldsymbol{\R}^{(b)} \left(\w(t)\right) + K_{\tau}\boldsymbol{\R}(t) \}$. The distribution of the elements of $C_{ij}$ for the white noise and for the weighted sum are indistinguishable (Fig. ~\ref{fig:validation}b), meaning that the components of the total noise used during integration are uncorrelated.

        Moreover, the mini--batch noise is not self--correlated in time, which is another necessary property of the noise used within a Langevin scheme (Fig.~\ref{fig:validation}c). In fact, the autocorrelation
        \begin{equation*}
            \overline{\chi}\left(\Delta t\right) = N^{-1} \sum_{i=1}^{N} \langle w_i(t+\Delta t)\,w_i(t) \rangle - \langle w_i(t+\Delta t) \rangle \langle w_i(t) \rangle,
        \end{equation*}
        of the mini-batch noise as a function of the time interval $\Delta t$, averaged over the components of the weight vector and over the starting time $t$, drops to zero essentially at the first time step (Fig.~\ref{fig:validation}c).

        Finally, we verified that the probability distribution sampled by the pL algorithm is compatible with that of the hMC, assumed as ground truth. For this purpose, we equilibrated each of the trained vectors at different temperatures using both sampling methods, and we measured at each temperature the mean and the standard deviation for the loss function, for the squared norm of the weights and for the training error $\epsilon$ as predicted by both algorithms.

        All the statistical properties of the model are very similar to each other (blue and orange curves in Fig.~\ref{fig:validation}d). We also calculated the relative error with respect to the hMC prediction, that is $|\langle O \rangle_T^{pL} - \langle O \rangle_T^{hMC}|/\langle O \rangle_T^{hMC}$, for each observable $O$ and temperature $T$. The prediction error is always of the order $\approx2 \cdot 10^{-2}$ and increases only slightly at low temperatures (with respect to the mean loss function), where the system can get trapped in local minima, making the equilibration of the system problematic and the results dependent on the specific simulations.

    \subsection{Better performance of the pL algorithm}
        \label{sec:performance}
        \begin{figure}
    		\centerline{\includegraphics[width=0.9\linewidth]{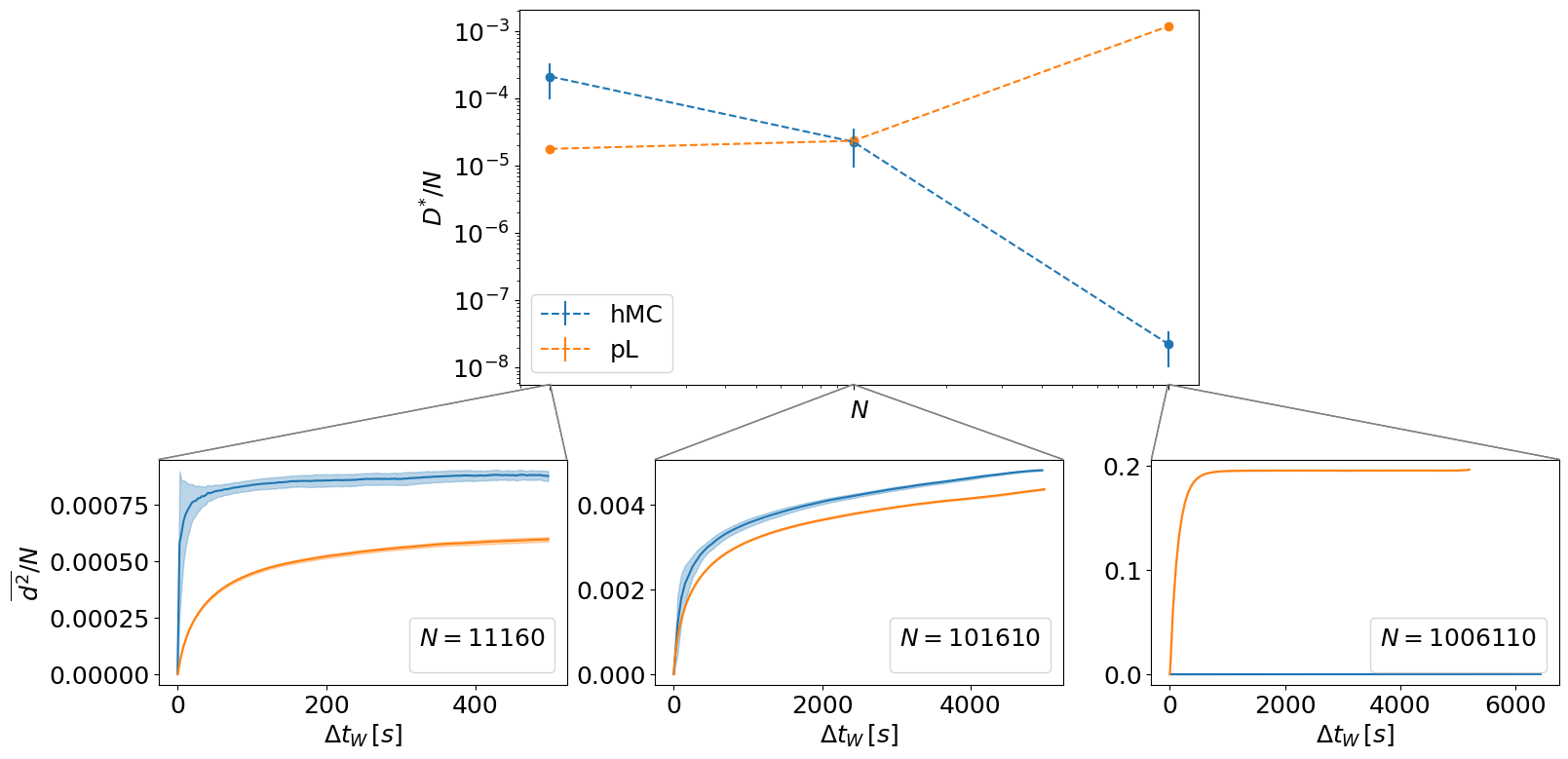}}
    		\caption{
            In the upper plot, the diffusion coefficient $D^*$ scaled by the network dimension $N$ as a function of $N$, for the three studied sizes $N=11160$, $N=101610$ and $N=1006110$ and for both sampling methods hMC (blue curve) and pL (orange curve).
            In the lower plots, the mean squared distance $\overline{d^{2}}$ scaled by the network dimension $N$ as a function of wall-clock time interval $\Delta t_W$ between sampled vectors, for the three studied sizes $N=11160$, $N=101610$ and $N=1006110$ and for both sampling methods hMC (blue curve) and pL (orange curve).
            The mean squared distances are obtained by averaging the values from independent simulations starting from different equilibrated weight vectors.
            }
    		\label{fig:performance}
    	\end{figure}

        We compared the performance of the pL scheme with that of the hMC for networks of different sizes (see Sect.~\ref{sec:results}). Performance was quantified by measuring the diffusion of the weight vector at equilibrium in wall-clock time and the hyper--parameters of both models were optimized to maximize diffusion while maintaining the correctness of the sampled distribution (see Fig. \ref{app_fig:prediction_comparison}). The simulations were performed on a dedicated 24-core Intel Xeon W5-3423 processor equipped with a Nvidia RTX 4500 Ada graphics card for this test only.
        
        Operatively, we performed independent samplings using both schemes. For the hMC method, we scanned the integration time step $\delta t$ and the total integration time $\Delta t$ per MC acceptance step, whereas for the pL algorithm, we tested different values for the initial temperatures ratios $\mathcal{T}_i$ and for the viscosity term $c_1$, while the size of the mini--batches was kept fixed at $S = 10^{-2}P$ for each model (see Fig.\ref{app_fig:lower_S} for pL performance with lower mini--batch sizes). All simulations were performed in triplicate at low temperature ($T=10^{-6}$), at which the system learns the input data well (cf. Fig. \ref{fig:validation}d and lower plot of Fig. \ref{fig:generalization}), starting from pre--equilibrated weights. For each sampling algorithm and for each model size $N$, we performed several simulations with different values of the hyper--parameters and selected those that maximize $\overline{d^2}\left( \Delta t_W\right)$, that is the mean squared distance computed between all the pairs of weight vectors separated by a wall--clock time $\Delta t_W$ (see Fig. \ref{app_fig:suboptimal_performance} for examples of sub--optimal parameters for the hMC and pL methods).

        The hybrid Monte Carlo scheme shows an edge for small network sizes (Fig.~\ref{fig:performance}, lower left and lower center panels). This is mainly because for small datasets the calculation of the full-batch gradient can be run in parallel, strongly reducing the advantage of calculating the gradient on a mini-batch of size $S \ll P$. However, for increasing $N$, the pL scheme clearly takes the lead (Fig.~\ref{fig:performance}, lower right plot), reaching the equilibrium distance while the hMC scheme cannot move appreciably from its starting point. 

        The computational efficiency of the two algorithms can be quantified by a "computational diffusion coefficient" $D^*$, which can be defined as the "derivative" of the $\overline{d^2}/N$ curve for $\Delta t_W \rightarrow 0$. Operatively, we measure
        \begin{equation*}
            D^* = \frac{\overline{d^2}(\Delta t^{\mathrm{min}}_W) - \overline{d^2}(0)}{\Delta t_W^{\mathrm{min}}}
        \end{equation*}
        where $\Delta t_W^{\mathrm{min}}$ is the minimum wall-clock time interval between sampled vectors (see Table \ref{app_tab:DtW_min} in the Appendix). As we can see in the upper panel of Fig.~\ref{fig:performance}, for hMC the relative computational diffusion coefficient $D^*/N$ drops rapidly as a function of the model size $N$, while it increases for pL. 
        
        These data suggest that the hMC scheme is basically useless for networks of realistic size, while pL can be used efficiently to sample their space of parameters.

    \subsection{Sampling with pL algorithm at intermediate temperatures provides optimal generalization properties}
        \label{sec:generalization}
        \begin{figure}
    		\centerline{\includegraphics[width=0.5\linewidth]{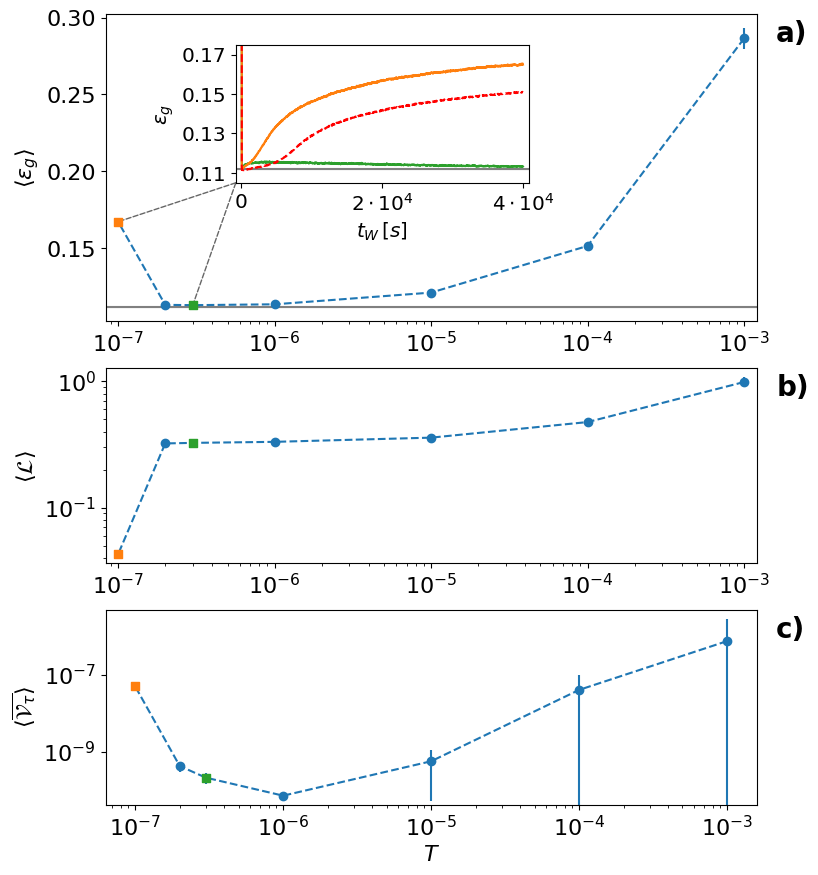}}
    		\caption{
            The mean value and the standard deviation of the generalization error $\epsilon_{g}$ \textbf{(a)}, the mean cross-entropy function $\mathcal{L}$ \textbf{(b)} and the average of all the variances of the mini--batch gradient components in the first two layers $\overline{\mathcal{V}_\tau}$ \textbf{(c)} sampled at equilibrium at different temperatures $T$ using the pL scheme.
            In the inserted plot, the generalization error $\epsilon_{g}$ as function of the wall-clock time $t_W$ during two simulations starting from initialized models at two different temperatures, $T=1.0\cdot10^{-7}$ (orange curve) and $T=3.0\cdot10^{-7}$ (green curve), and during an Adam training beyond early--stopping (red dotted curve). 
            The gray straight lines reported in the upper plot and in the inserted one represent the best mean generalization error found with Adam training.
            All values of $\epsilon_g$ have been computed on the same test dataset.
            }
    		\label{fig:generalization}
    	\end{figure}

        We investigated the generalization properties of the weights sampled with the pL algorithm at different temperatures, comparing them with those obtained from a SGD. We focused our attention to the largest network ($N=1006110$), since it is the most realistic size in practical applications.

        We calculated at different temperatures the average of the generalization error $\epsilon_g$ [cf. Eq. \eqref{eq:errorfunction}] on a test set of size $P_{g} \approx 0.18P$,  generated as described in Sect. \ref{sec:model_and_task} using the same spin-flip probability and the same reference vectors as in the training set.

        We observe that the average loss $\langle\mathcal{L}\rangle$ displays a sudden increase, reminiscent of a first-order phase transition, at $T_f \approx 2\cdot 10^{-7}$ (Fig. \ref{fig:generalization}b). Interestingly, this transition seems to affect also the generalization properties of the sampled vectors, as the average generalization error is larger at very low temperatures and displays a sudden drop at the critical temperature at which the average loss increases (Fig. \ref{fig:generalization}a). The average generalization error then increases, assuming its minimum value just above the transition temperature.

        The minimal value $\langle\epsilon_g\rangle\approx 0.113$ is very close to the lower bound that the generalization error can assume based on geometric arguments. As anticipated, for $p_f=0.355$, the probability that a spin vector of class $\mu$ is closer to the reference vector of another class $\nu$ than to its own is approximately $\approx 0.1$, which implies that approximately a tenth of the spin vectors do not display the properties that define their own class (and thus can't be classified correctly without explicit training on such data).
        
        In the low--temperature phase, the system loses some of its generalization capabilities. For example, at $T=10^{-7}$, the average prediction error in the training set is $\langle \epsilon \rangle \approx 0.014$ while it is $\langle \epsilon_g \rangle \approx 0.17$ in the test set (Fig. \ref{fig:generalization}a) suggesting that at these temperatures the system overfits the training data.

        We also compared the results from the samplings with those obtained by training three independent models by SGD using the Adam algorithm (with a learning rate $\eta = 10^{-5}$) and early--stopping procedure (on a validation subset of the training data with size $P_v \approx 0.18P$). Starting from random weights, both technique minimize their error on the test set very rapidly (inset of Fig. \ref{fig:generalization}a), as both descents are of the order of seconds. The best models found through standard training techniques have a mean generalization error $\langle \epsilon_{g} \rangle_{\mathrm{Adam}} \approx 0.112$, but the latter increases markedly afterwards (inset of Fig. \ref{fig:generalization}a, red dotted curve) if early--stopping is not applied. Also, the optimal generalization error found by the Adam algorithm is similar to that of the pL sampling at intermediate temperature. However, the pL scheme does not require an early--stopping procedure (and thus a validation set) to avoid overfitting, as the generalization properties of the sampled weights are uniquely determined by the temperature $T$.
       
        In fact, during the training (inset of Fig. \ref{fig:generalization}a), the low--temperature ($T=1.0 \cdot 10^{-7}$) simulation starting from random initialized vectors reaches a minimum of generalization error and rapidly departs from it, as the Adam algorithm does. Vice versa, during the sampling at intermediate temperature ($T=3.0 \cdot 10^{-7}$), the system fluctuates steadily around the low equilibrium value without overfitting.

        Lastly, we observe a correlation between the generalization error $\langle \epsilon_g \rangle$ and the average of all the variances associated to weights in the first two layers $\langle \overline{\mathcal{V}_\tau} \rangle$ (which account for $99.4\%$ of the network weights) as functions of the temperature $T$ (Fig. \ref{fig:generalization}a and Fig. \ref{fig:generalization}c). The minimum values for each observable are reached in a temperature interval going from $T = 2 \cdot 10^{-7}$ (right above the phase transition) up to $T = 10^{-6}$, where the generalization error has a flat shape at approximately $\langle \epsilon_g \rangle \approx 0.113$.


    \subsection{Test on realistic data}
        \label{sec:generalization}
        \begin{figure}
    		\centerline{\includegraphics[width=0.7\linewidth]{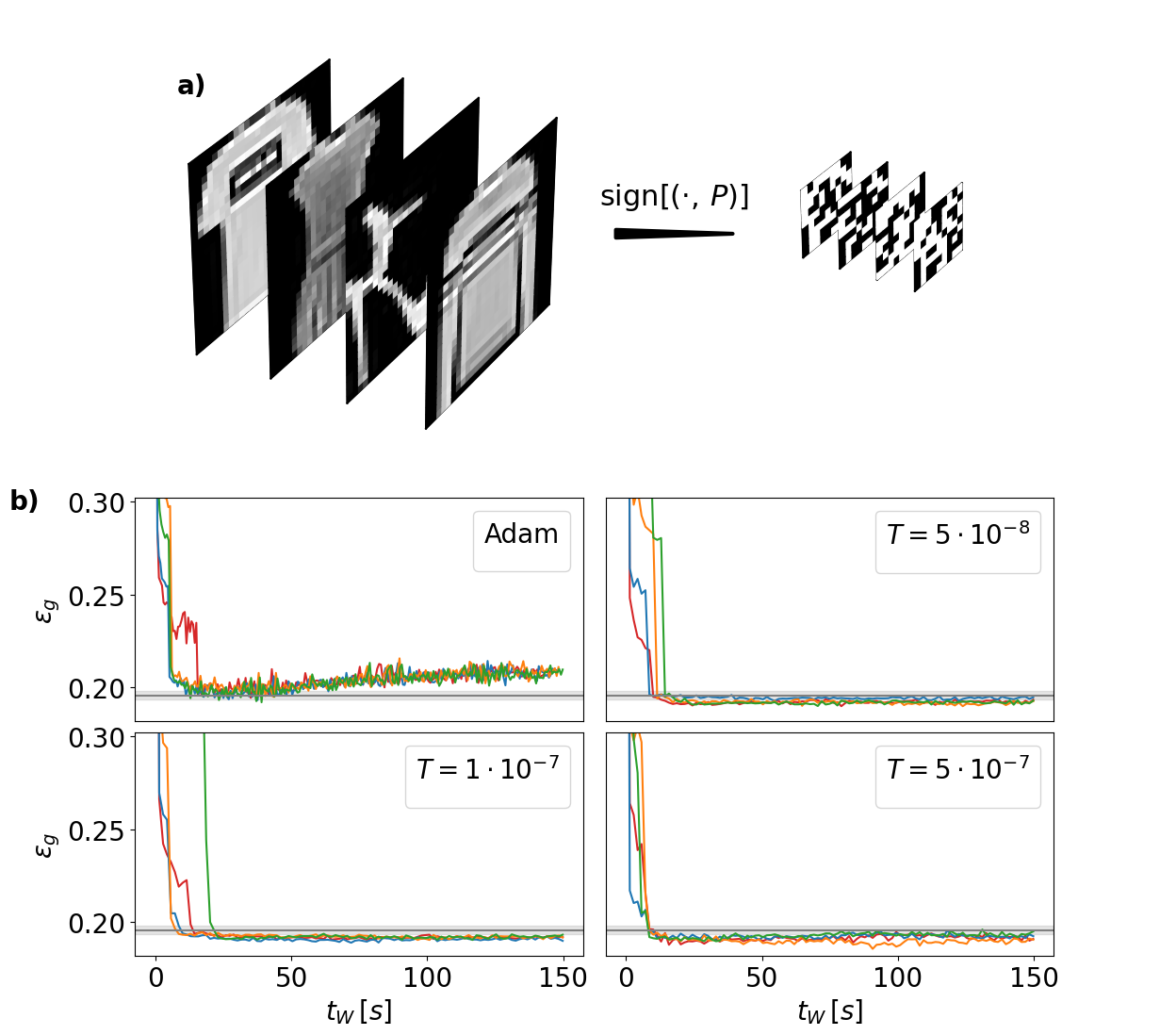}}
    		\caption{
            \textbf{a)} Sample images $\left( 28\times28 \right)$ of the FashionMNIST training set and their projection into $d=100$ binary vectors used during training and pL samplings. Each $d$-vector is represented as a $\left( 10\times10 \right)$ matrix. 
            \textbf{b)} The generalization error $\epsilon_g$ computed on the test set of the FashionMNIST dataset as a function of the wall--clock time $t_W$ during trainings with Adam as optimizer (upper left) and pL samplings performed at different temperatures, i.e. $T=5 \cdot 10^{-8}$ (upper right), $T=1 \cdot 10^{-7}$ (lower left) and $T=5 \cdot 10^{-7}$ (lower right). Curves with the same color in different plots start from the same stochastically generated weight vector.
            }
    		\label{fig:FMNIST}
    	\end{figure}

        We tested our sampling method on the FashionMNIST dataset \cite{Xiao2017Fashion-MNIST:Algorithms}, composed of $\left( 28\times28 \right)$ images ($P=60000$ for training and validation, and $P_g=10000$ for evaluation) and $K=10$ classes.
        To keep the same architecture used for the spin vectors case, we adapted the input data, by generating a matrix $P \sim \{ \pm 1 \}^{28\times28\times d}$  and projecting each image into a $d$ vector as 
        \begin{equation}
            \tilde{x}^{\mu}_k = \mathrm{sign} \left[ \sum_{ij} x^{\mu}_{ij} \cdot P_{ijk}\right]
            \label{eq:projectedFMNIST}
        \end{equation}
        (cf. Fig. \ref{fig:FMNIST}a). We performed several pL samplings  ($T \in \left[ 10^{-7}, 10^{-6} \right]$ and $\lambda=10^{3}$) from stochastically generated weight vectors. For each simulation, we also trained the model using the Adam optimizer with the optimal learning rate $\eta =10^{-4}$ and the same $L_2$ regularization coefficient as for pL. 

        We observe that the two methods are basically equivalent in terms of time--efficiency, as both require $t_W \approx 30 \,s$ to reach the minimum value in generalization error (Fig. \ref{fig:FMNIST}b). However, as in the case of binary spin vectors (Sect. \ref{sec:generalization}), during the pL samplings, the test error fluctuates around its equilibrium value of $\langle \epsilon_g \rangle \approx 0.2$ for every temperature in the intermediate range $T \in \left[ 5 \cdot 10^{-8}, 5 \cdot 10^{-7} \right]$ [Fig. \ref{fig:FMNIST}b, upper right and lower plots]. On the contrary, during each Adam training the model quickly starts overfitting soon after the optimal generalization error has been reached [Fig. \ref{fig:FMNIST}b, upper left plot].

\section{Discussion and Conclusions}

    Sampling the parameter space of ANN can be useful in many ways. As well as providing insight into the topology of the solution space and how the system functions, sampling can be used to obtain sets of parameters with good predictive capabilities. We have demonstrated that sampling the Boltzmann distribution at intermediate temperatures yields sets of parameters with optimal generalization error, eliminating the need for early stopping procedures and the usage of a portion of the training dataset for validation.

    Also Bayesian networks have the tendency to remove overfitting \cite{Welling2011BayesianDynamics}. The sampling scheme we propose has the advantage of not requiring a complete reconstruction of the posterior probability $p(\w|\mathcal{D})$ if one is interested only in a prediction and not in the associated probability. Since the fluctuations in loss and in generalization error are very small at intermediate temperatures, any set of parameters selected randomly at equilibrium is equally good for predicting the output.

    One possible problem when using pL as a training tool is the need to select 'intermediate' temperatures. In the low-temperature regime, the system overfits in a manner similar to that observed in gradient-descent minimization. At high temperatures, however, it fails to learn the training dataset. However, at least in the cases analyzed (that is, spin vectors and projected FashionMNIST \cite{Xiao2017Fashion-MNIST:Algorithms}), the range of 'intermediate' temperatures spans several orders of magnitude, therefore should be easily identified, especially using the variance of the gradient over mini--batches as a fast order parameter. 
    
    The use of mini--batches within the pL scheme permits the sampling of the space of parameters of large networks, usually trained with a large amount of data. The variance of the gradient over mini--batches appears as a powerful proxy for the ability of the network to generalize, and thus to identify the correct temperatures on the fly. The rationale for the correlation between gradient variance and generalization error can be found in the flatness of the loss minima. As shown in \cite{Baldassi2015}, wider minima have better generalization properties than narrower ones, even with similar loss. On the other hand, we expect the energy profile of each stochastic mini-batch in flat minima to share similar properties across mini-batches, including the gradient (as shown in ref. \cite{Feng2021TheMinima}). Therefore, the variance of the mini--batch gradient is a useful quantity that can be monitored to conveniently study the hyperparameters of the network.

    The pL seems the only algorithm suitable for sampling a large ANN according to Boltzmann probability. Monte Carlo techniques guarantee that the sampling converges to the desired distribution, but usually suffer the high dimensionality of the space, where random moves tend to lead to high--energy conformations, thus reducing drastically the acceptance probability. The hybrid Monte Carlo of ref. \cite{Zambon2025SamplingNetwork} amends this problem, proposing moves that are guided by the gradient and still obeys the principle of detailed balance. However, the exact calculation of the gradient is computationally demanding and limits the application of the hMC to systems whose training dataset is much smaller than those used in everyday applications. While loss minimization techniques solve the problem by using mini--batches, in hMC the estimation of the gradient on small subsets of the training data results in a negligibly small acceptance probability.
   
    The pL algorithm combines the ability to sample asymptotically the correct distribution with the computational efficiency given by the use of mini--batches. We have shown that the system can diffuse efficiently in the manyfold of low--loss parameters, even for large systems trained by massive datasets. A key observation which allows the algorithm to work is that the noise induced by the approximate estimation of gradients is Gaussian for mini--batches of nontrivial numerosity, thanks to the central limit theorem. Moreover, we can control the effect of this noise in the Langevin equations by an optimal choice of the fictitious masses and the fictitious friction coefficient associated with the parameters of the system, even without using a negligible timestep, in order to sample the Boltzmann weight associated with the desired temperature.

    Overall, our results suggest that controlled mini--batch Langevin dynamics provides a principled and scalable route to explore and exploit the low-loss manifold of large ANN, replacing optimization with equilibrium statistical mechanics.

\newpage
\bibliography{references}

\newpage

\appendix
\section*{Appendix}
\setcounter{figure}{0}            
\renewcommand\thefigure{A.\arabic{figure}} 
\renewcommand{\thetable}{A\arabic{table}}

    \begin{table}[h]
        \centering
        \begin{tabular}{c|c|c}
            \hline
              $N$        & hMC $[s]$        & pL $[s]$                              \\
              \hline
              $11160$    & $2.7 \cdot 10^0$ & $\left(2.0+1.5\right) \cdot 10^{-3}$  \\ 
              $10160$    & $5.1 \cdot 10^1$ & $\left(3.5+3.2\right) \cdot 10^{-3}$  \\ 
              $1006110$  & $3.2 \cdot 10^3$ & $\left(5.2+5.1\right) \cdot 10^{-2}$  \\
            \hline
        \end{tabular}
        \caption{
        The minimum wall-clock time interval between sampled vectors $\Delta t_W^{\mathrm{min}}$ for all model sizes $N$, and for both sampling schemes, hMC and pL. For hMC, we report the times relative to a single move with the optimal number of integration steps, that is $10^3$ for all model sizes. For pL, the first value within the parenthesis represents the time needed for a single integration, while the second one is the time spent to extract $100$ mini--batches for the estimate of $\mathcal{V}_{\tau}$. Also, all pL values are relative to a mini--batch size $S=10^{-2}P$.
        }
        \label{app_tab:DtW_min}
    \end{table}

    \begin{figure}[h]
    	\centerline{\includegraphics[width=0.715\linewidth]{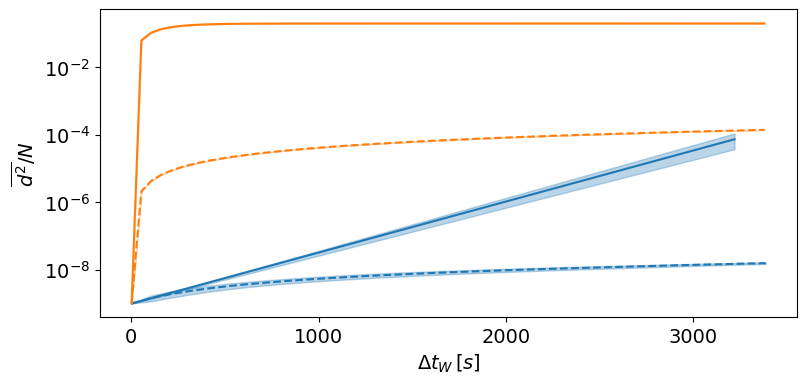}}
    	\caption{
        The normalized mean squared distance $\overline{d^{2}}/N$  as a function of wall-clock time interval $\Delta t_W$, for the biggest model size $N=1006110$ sampled with hMC (thick blue curve), mb-hMC (dotted blue curve), pL (thick orange curve) and naive-pL (dotted orange curve).
        The mean squared distances are obtained by averaging the values from independent simulations starting from different equilibrated weight vectors.
        All curves have been shifted so that $\langle d^{2} \rangle(0) = 10^{-9}$, since the y-axis is in log-scale.
        }
    	\label{app_fig:othermethods_performance}
    \end{figure}

    \begin{figure}[h]
    	\centerline{\includegraphics[width=0.71\linewidth]{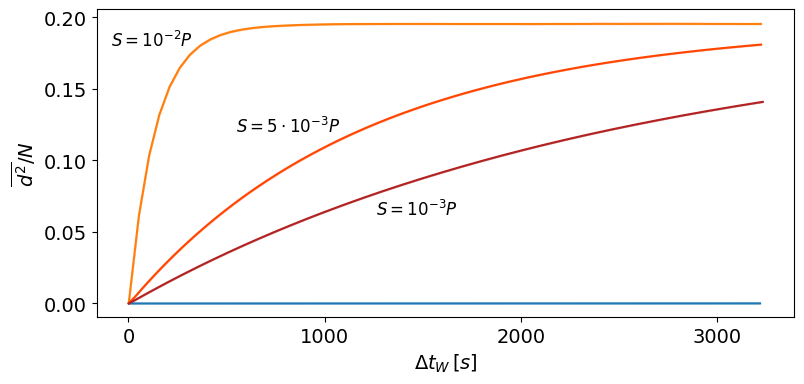}}
    	\caption{
        The normalized mean squared distance $\overline{d^{2}}/N$  as a function of wall-clock time interval $\Delta t_W$, for the biggest model size $N=1006110$ using pL with $S=10^{-2}P$ (upper orange curve), pL with $S=5\cdot10^{-3}P$ (intermediate dark-orange curve), pL with $S=10^{-3}P$ (lower red curve) and hMC (blue curve).
        The mean squared distances are obtained by averaging the values from independent simulations starting from different equilibrated weight vectors.
        }
    	\label{app_fig:lower_S}
    \end{figure}

    \begin{figure}
    	\centerline{\includegraphics[width=0.7\linewidth]{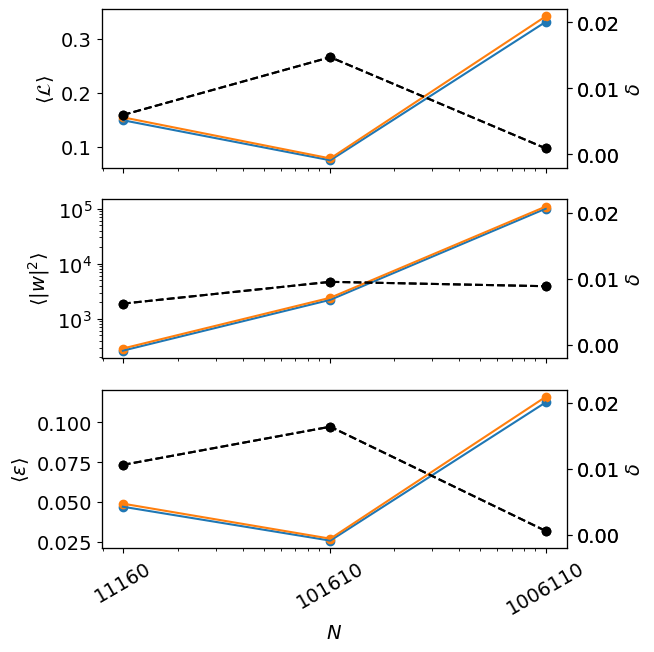}}
    	\caption{
        The average of the mean cross-entropy $\mathcal{L}$ (upper plot), the squared norm $|\w|^2$ (center plot) and the training error $\epsilon$ (lower plot) as a function of the size $N$ of the model at $T=10^{-6}$, obtained from hMC (blue curves) and pL (orange curves) simulations at equilibrium. The orange and blue curves have been slightly shifted along the y-axis to make them distinguishable.
        The relative error with respect to the hMC prediction between the estimates is also reported (gray dotted line).}
    	\label{app_fig:prediction_comparison}
    \end{figure}

    \begin{figure}
    	\centerline{\includegraphics[width=0.8\linewidth]{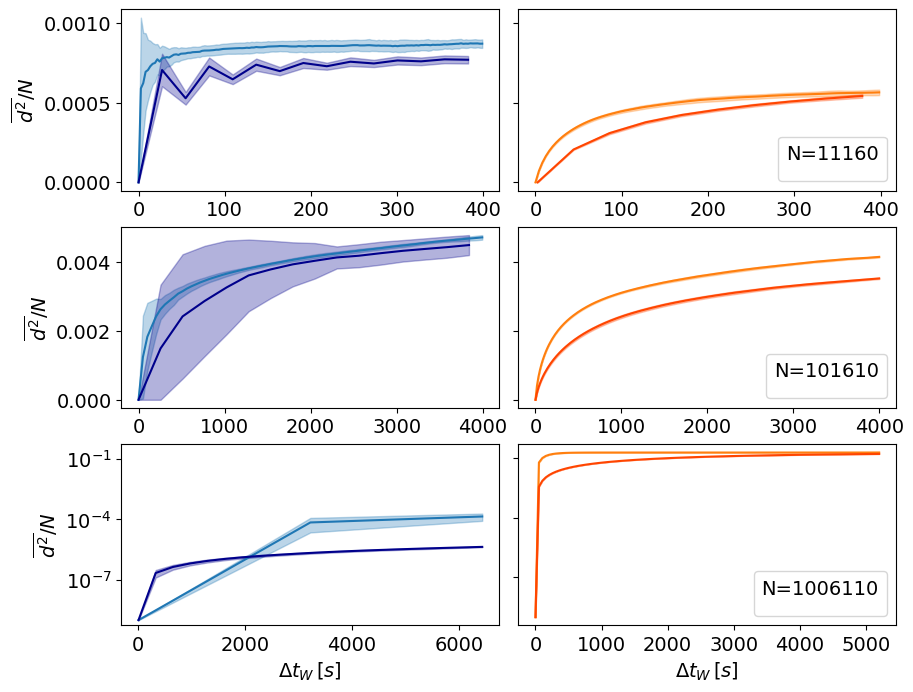}}
    	\caption{
        The relative mean squared distance $\overline{d^{2}}/N$ as a function of wall-clock time interval $\Delta t_W$ between sampled vectors, for the three studied sizes $N=11160$, $N=101610$ and $N=1006110$  and for both hMC and pL (left and right columns, respectively). For each size and for each method, the optimal mean squared distance curve (brighter color) is shown along with a suboptimal one (darker color). 
        }
    	\label{app_fig:suboptimal_performance}
    \end{figure}

    \begin{figure}
    	\centerline{\includegraphics[width=0.7\linewidth]{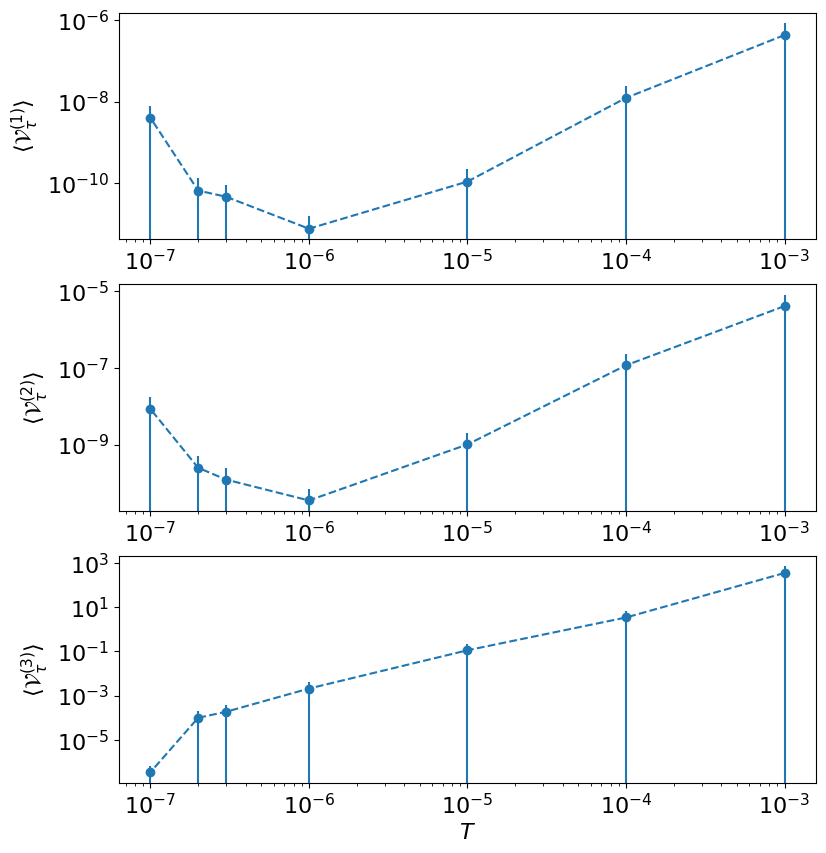}}
    	\caption{
        Mean and standard deviation of the average of all the variances in the first layer (upper plot), in the second layer (middle plot) and in the third layer (lower plot) for the weight vectors sampled at equilibrium at different temperatures $T$ using the pL scheme.
        }
    	\label{app_fig:suboptimal_performance}
    \end{figure}

\end{document}